\documentclass[mathleft
]{an}
\usepackage{graphicx}
\usepackage{times}
\pdfoutput=1
\begin{document}

\Pagespan{1}{6}
\Yearpublication{2012}%
\Yearsubmission{2012}%
\Month{??}%
\Volume{??}%
\Issue{??}%

\title{Seismology of active stars}

\author{S. Hekker\inst{1}\fnmsep\thanks{Corresponding author:
  \email{S.Hekker@uva.nl}\newline}
\and  R.A. Garc\'ia\inst{2}
}
\titlerunning{Seismology of active stars}
\authorrunning{S. Hekker \& R.A. Garc\'ia}
\institute{
Astronomical Institute 'Anton Pannekoek', University of Amsterdam, Science Park 904, 1098 XH Amsterdam, the Netherlands
\and 
Laboratoire AIM, CEA/DSM-CNRS, Universit\'{e} Paris 7 Diderot, IRFU/SAp, Centre de Saclay, 91191, GIf-sur-Yvette, France}

\received{15 August 2012}
\accepted{??}
\publonline{??}

\keywords{stars: oscillations -- stars: activity --stars: atmospheres -- stars: interiors}

\abstract{In this review we will discuss the current standing and open questions of seismology in active stars. With the longer photometric timeseries data that are --and will become-- available from space-missions such as \textit{Kepler} we foresee significant progress in our understanding of stellar internal structures and processes --including interactions between them-- taking place in active stars in the next few years.}

\maketitle

\section{Introduction}
To fully understand the internal structures of stars and the structure changes over time, all processes defining the stellar properties and the interactions between these processes need to be investigated. Among these processes are: fusion reactions in the core, diffusion / mixing of chemical elements, radiative and convective transport, overshooting, rotation, ionisation, magnetic fields, oscillations, mass-loss / flares / winds from the surface and so on. For some of these processes direct observational signatures can be measured, while others have to be inferred in an indirect manner.

Here we review oscillation processes in active solar-like stars, where we use a working definition of solar activity in the context of stars as proposed by Judge \& Thompson 2012: `globally-observable variations on time scales below thermal time scales of 10$^5$ years for the convection zone'. Following this definition, activity is dominated by magnetic-field evolution. 

In the Sun, the magnetic evolution is observed as a Hale cycle: this is an approximately 22-year cycle in which the magnetic polarity of sunspot pairs reverses and then returns to its original state; during half the cycle, the leading spot in every pair will have a positive polarity but during the other half, the leading spot will be of negative polarity. Oscillatory dynamo models explaining this cycle will be discussed in Section 2.

In this work we will discuss the interaction between activity and solar-like oscillations. Solar-like oscillations are acoustic waves stochastically excited  in the convective atmospheres of solar-like main-sequence stars, subgiants and red giants. For these stars, only low-degree ($\ell =0,1, 2, 3$) modes can be observed (see Section 3 for more details). From Sun-as-a-star observations it has been shown that the oscillation signature changes in correlation with surface magnetic-activity indicators (e.g., Jim\'enez Reyes et al. 2004; Jain et al. 2009; Broomhall et al. 2011a). This correlation will be discussed in Section 4. In Section 5 we describe parallels between results obtained for the Sun and other stars. We finish with some future perspectives.

\section{The Hale cycle}
\begin{figure}
\includegraphics[width=\linewidth]{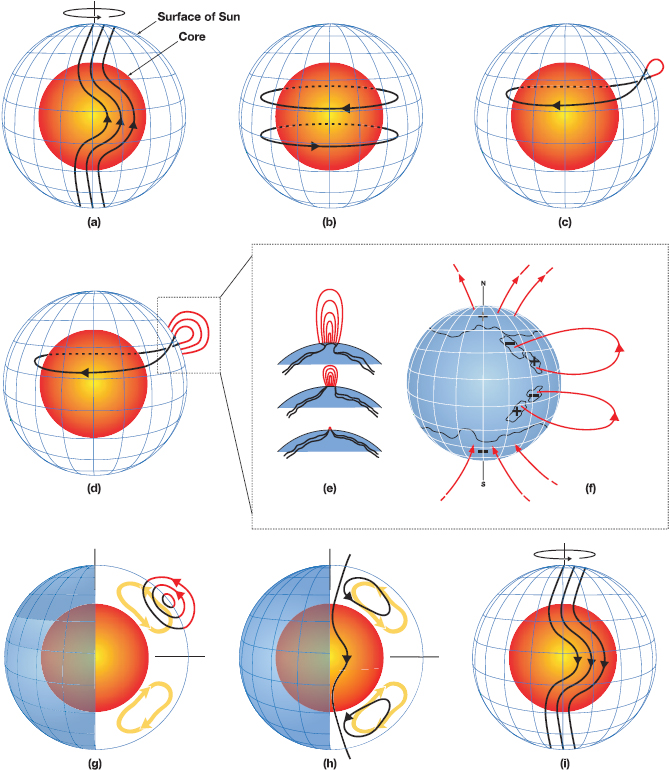}
\caption{Schematic of solar flux-transport dynamo processes. Red inner sphere represents the Sun's radiative interior and blue mesh the solar surface. In between is the solar convection zone where dynamo resides. (a) Shearing of poloidal field by the Sun's differential rotation near convection zone bottom. The Sun rotates faster at the equator than the pole. (b) Toroidal field produced due to this shearing by differential rotation. (c) When toroidal field is strong enough, buoyant loops rise to the surface, twisting as they rise due to rotational influence. Sunspots (two black dots) are formed from these loops. (d--f) Additional flux emerges (d, e) and spreads (f) in latitude and longitude from decaying spots (as described in figure 5 of Babcock 1961). (g) Meridional flow (yellow circulation with arrows) carries surface magnetic flux poleward, causing polar fields to reverse. (h) Some of this flux is then transported downward to the bottom and towards the equator. These poloidal fields have sign opposite to those at the beginning of the sequence, in (a). (i) This reversed poloidal flux is then sheared again near the bottom by the differential rotation to produce the new toroidal field opposite in sign to that shown in (b). Figure taken from Dikpati \& Gilman (2007).}
\label{hale}
\end{figure}

The 22-year Hale cycle (quasi-periodic reversal of the global magnetic field in the Sun) can be described by ($\alpha \Omega$) dynamo models (see e.g., Rempel 2008; Charbonneau 2010 and references therein). The basic processes involved in these models are the generation of toroidal field by shearing a pre-existing poloidal field by differential rotation ($\Omega$-effect) and the re-generation of poloidal field by lifting and twisting the toroidal flux tube by helical motion ($\alpha$-effect).  Over the years different large-scale $\alpha\Omega$ dynamos have been developed: a convection-dynamo with the $\alpha$ and $\Omega$-effect working in the bulk of the convection zone (e.g., Durney \& Latour 1978), a thin-layer dynamo with the $\alpha$ and $\Omega$-effect working in the tachocline (e.g., Schmitt \& Sch\"ussler 1989), an interface dynamo with the $\alpha$ and $\Omega$-effect working at the base of the convection zone with a slight separation between them (e.g., Charbonneau \& MacGregor 1997), a pure Babcock-Leighton dynamo with the $\alpha$-effect working at the surface and the $\Omega$-effect at the tachocline (Babcock 1961, Leighton 1964, 1969) and flux-transport dynamo which include the advective transport of magnetic flux by meridional circulation in addition to the $\alpha$ and $\Omega$-effect (e.g., Schmitt 1987, Choudhari et al. 1995, Ossendrijver 2000). The latter model seems to be the only one to be able to explain the equator-ward migration of spots, and also the pole-ward drifting large-scale diffuse fields (Dikpati 2005). 

The flux-tube model works as follows. Starting the cycle when a poloidal magnetic field is present. Differential stellar rotation with an equator to pole gradient will distort this poloidal field stretching it somewhere at the bottom of the convection zone (the $\Omega$-effect). The field lines are stretched further at locations with larger rotational velocities, i.e., closer to the equator. This will eventually result in a toroidal magnetic field. The differential rotation reinforces the toroidal field until it becomes in an unstable state, when the buoyant force begins to act on it. Then, the toroidal magnetic field re-arranges into vast flux tubes which start to twist and lift and emerge at the surface (the $\alpha$-effect). At this time, a typical magnetic feature appears: a magnetic loop, with a cool bipolar spot pair at the footpoints, or in other words: spots. The large scale meridional convection then acts on the distorted toroidal field and it regenerates the poloidal field. In this process the spots are transported towards the poles. This process results in a poloidal field, similar to that we had at the beginning, but the polarity is now the opposite in sign. 
This flux-tube dynamo is shown schematically in Fig.~\ref{hale}.

\begin{figure}
\includegraphics[width=\linewidth]{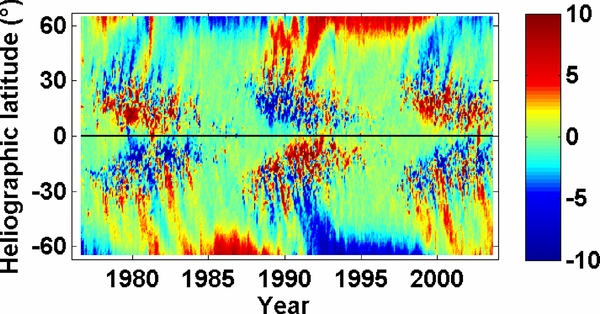}
\caption{Butterfly diagram of the net radial magnetic flux averaged over longitude for each Carrington Rotation (the Carrington coordinate system rotates rigidly with the mean solar rotation ratio (27.2753 days)). The magnetic flux in Gauss is colour-coded. The black horizontal line marks the lattitude $\theta = 0^{\circ}$. Figure taken from Vecchio et al. (2012), reproduced by permission of the AAS.}
\label{butterfly}
\end{figure}

The dynamo process can be observationally followed from the evolution of spots in the so-called 11 year cycle observed for the Sun. See Fig.~\ref{butterfly} for a representation in a 'butterfly diagram'. This diagram shows that the spots first appear at latitudes of about 35$^{\circ}$ and migrate towards the equator as the cycle evolves.

Understanding the formation, evolution and motion of star spots is of prime importance as this provides information on the distribution of surface magnetic field and with that of the dynamo processes in stars. As (differential) rotation is a key ingredient of the dynamo model, this has to be taken into account as well.

\section{Solar-like oscillations}
Solar-like oscillations are acoustic waves stochastically excited by convection in the turbulent outer layers of low-mass main-sequence stars, subgiants and red giants (see Aerts et al. 2010 for a more details on stellar oscillations). The waves resonate in a cavity spanning between an upper and a lower turning point. The upper turning point lies close to the surface and is defined by the cut-off frequency (e.g., Jim\'enez et al 2011). Above this turning point the atmosphere is not able to trap the modes and the oscillations become traveling waves, so-called high-frequency modes or pseudomodes (e.g., Jefferies et al. 1988; Garc\'\i a et al. 1998; Karoff et al. 2007). The location of the upper turning point depends on the frequency and wavenumber of the modes. The lower turning point lies at a depth inside the star at which the horizontal phase speed of the wave equals the local sound speed. The location of the lower turning point is mostly dependent on the degree of the modes.
\begin{figure}
\includegraphics[width=\linewidth]{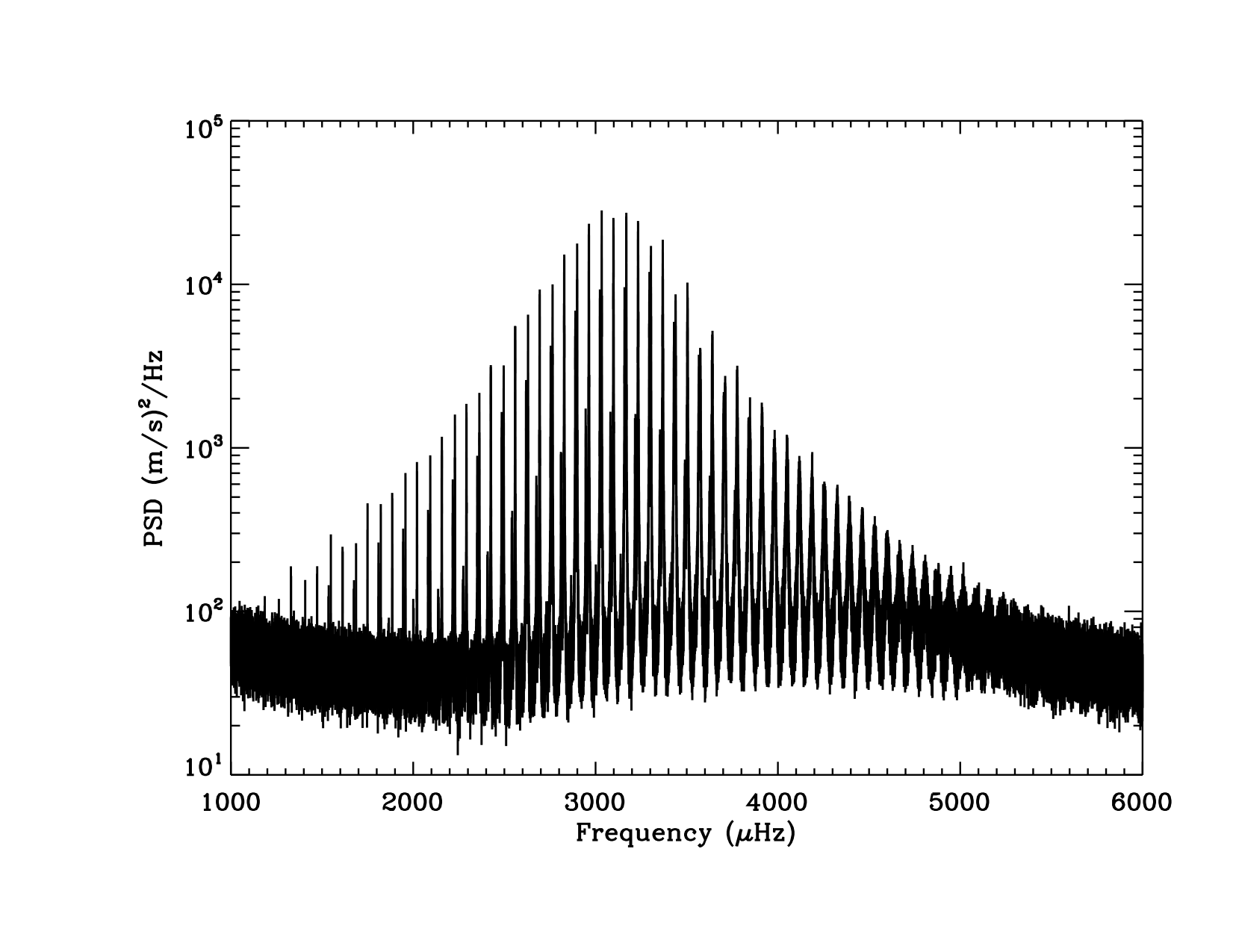}
\caption{Power spectrum density (PSD) of 14 years of GOLF velocity time series (Garc\'\i a et al. 2005) starting on April, 11, 1996. The PSD has been smoothed by a boxcar function of 67 nHz.}
\label{power spectrum}
\end{figure}
These oscillations present themselves as a nearly regular pattern in the Fourier spectrum of timeseries data (Fig.~\ref{power spectrum}). The locations of the frequencies with radial order $n$ and spherical degree $\ell$ ($\nu_{n,\ell}$) can be approximated by the asymptotic relation (Tassoul 1980):
\begin{equation}
\nu_{n,\ell}=\Delta\nu(n+\frac{\ell}{2}+\epsilon)+d_{0\ell}
\label{tassoul}
\end{equation}
in which $\Delta \nu$ is the nearly regular spacing between modes of the same degree and consecutive orders. This so-called large frequency spacing is proportional to the mean density of the star. $\epsilon$ is a phase term and $d_{0\ell}$ is the small frequency separation, a second order term dependent on mode degree. This relation has recently been updated by Mosser et al. (2011) to include the frequency dependence of $\Delta\nu$ (Hekker et al. 2011c).  From these oscillations it is possible to obtain `global' stellar properties such as mass, radius and mean density (e.g., Mathur et al. 2010; Chaplin et al. 2011b, Hekker et al. 2011a,b), as well as more detailed inferences such as the location of the second Helium ionization zone and the base of the convection zone (e.g., Basu \& Antia 1994, Monteiro et al. 1994).

The oscillations described here have pure acoustic pressure (p-)modes, which are observed in low-mass main-sequence stars. When stars evolve the p-mode frequencies decrease, while the frequencies of oscillations that are present in a cavity in the deep radiative interior of the star, so-called gravity (g-)modes with buoyancy as their restoring force, have increasing frequencies. At some stage a coupling between the cavities persists and a p- and g- mode with similar frequencies and same spherical degree undergo an avoided crossing (e.g., Bedding et al. 2011). This causes deviations from the near regular pattern as described above. These deviations are of particular interest to study the deeper interiors of the stars and to infer stellar ages with relatively high accuracy (e.g., Deheuvels \& Michel 2011). A connection between these so-called mixed modes (Osaki 1975; Aizenman et al. 1977) and activity is at present unknown and not further discussed here. 

\section{Oscillation - activity correlation}
\begin{figure*}
\begin{minipage}{0.495\linewidth}
\centering
\includegraphics[width=\linewidth]{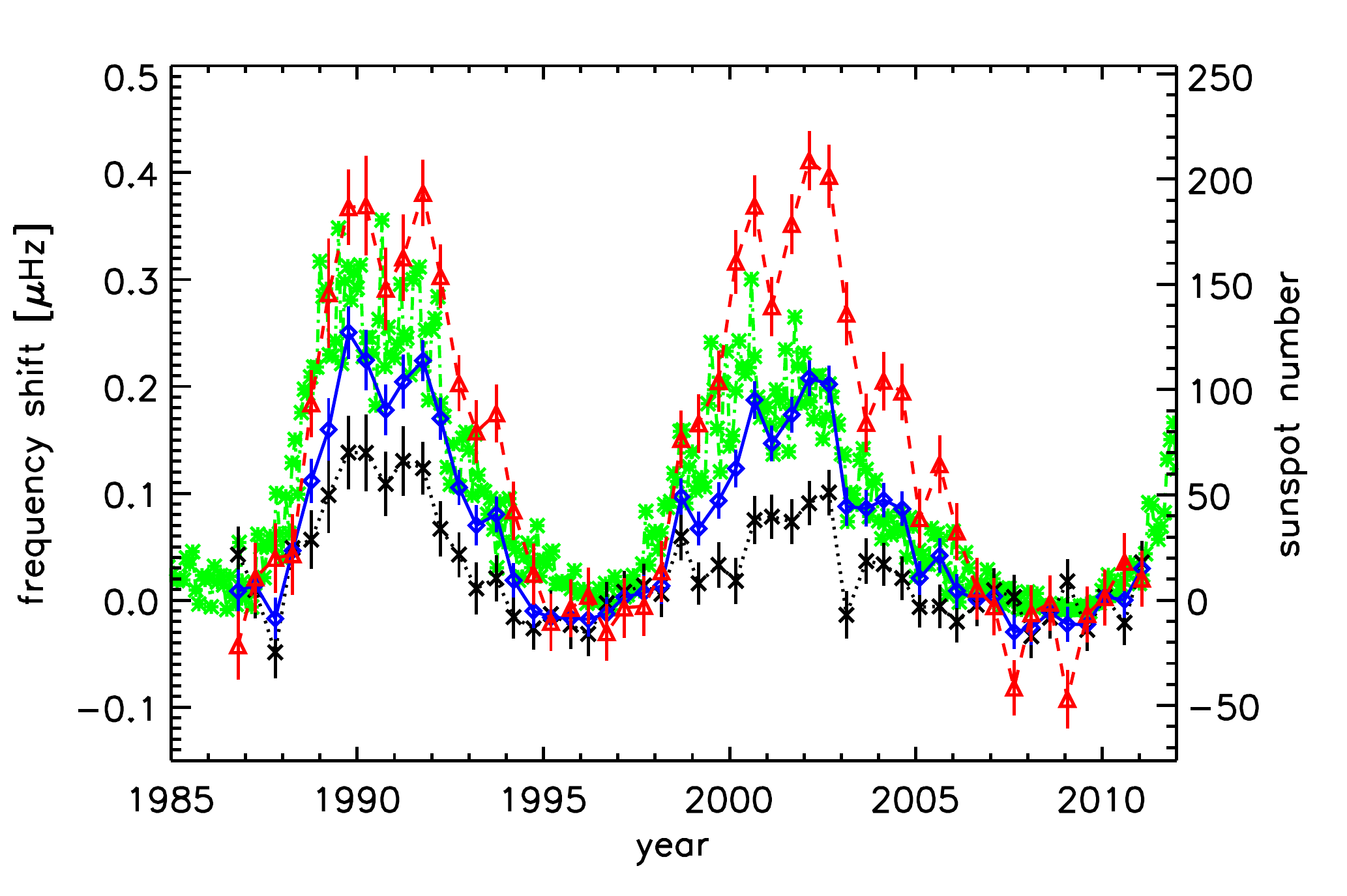}
\end{minipage}
\begin{minipage}{0.495\linewidth}
\centering
\includegraphics[width=\linewidth]{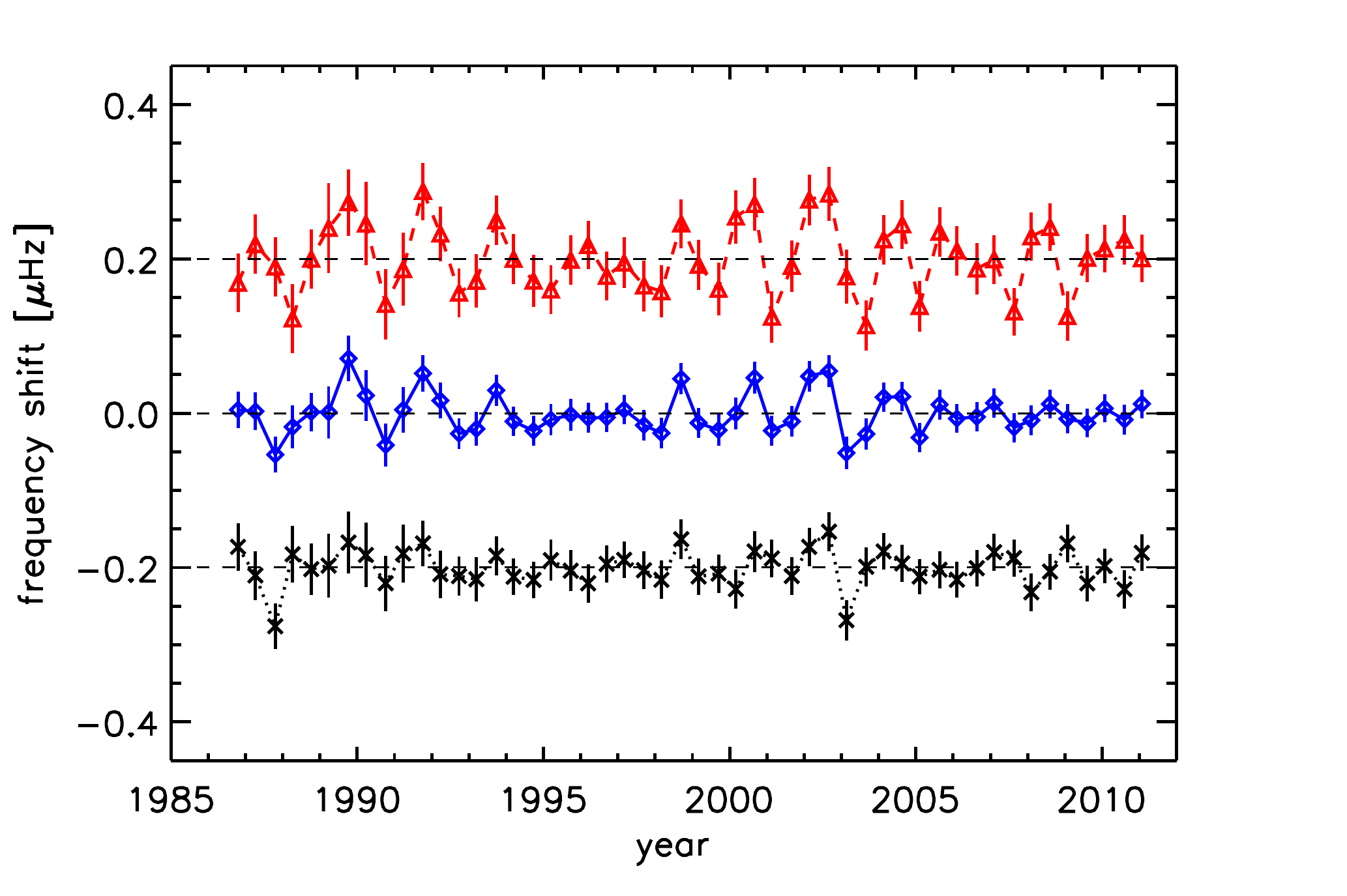}
\end{minipage}
\caption{Left: average frequency shifts of ÒSun-as-a-starÓ modes with frequencies between 1.88 and 3.71 mHz (total-frequency band, blue solid line, and diamond symbols); 1.88 and 2.77 mHz (low-frequency band, black dotted line, and cross symbols); and 2.82 and 3.71 mHz (high-frequency band, red dashed line, and triangle symbols). The international sunspot number (taken from National Geophysical Data Center, Solar and Terrestrial Physics: http://www.ngdc.noaa.gov/) is shown in green dashed-dotted line, and asterisks (right axis). Right: residuals left after dominant 11-year signal has been removed (dotted and red dashed curves are displaced by -0.2 and +0.2, respectively, for clarity). Figure based on BiSON data and adapted from Broomhall et al. (2011b).}
\label{cor}
\end{figure*}

Correlations between surface activity indicators and solar-like oscillations in the Sun were established in the early nineties (e.g., Elsworth et al. 1990; Libbrecht \& Woodard 1990), when dedicated solar instruments, such as BiSON (Birmingham Solar-Oscillations Network, Chaplin et al. 1996) and Big Bear Solar observatory (Zirin 1970, Denker et al. 2006) were long enough in place to have data obtained during a solar minimum and maximum (see Fig.~\ref{cor}). At the time these results were already expected as it was predicted that the magnetic fields, responsible for spots, induce perturbations in the outer parts of the star, which influence the oscillation cavity of the  p modes and thus the properties of the modes that probe this cavity. Indeed, in the Sun the oscillation frequencies are shifted to higher frequencies during higher activity. The exact mechanism behind this shift is however not yet known. Direct action of the magnetic fields on the tachocline, the sunspot anchoring zone, the photosphere or chromosphere does not seem to reproduce the observations. Other possibilities are indirect effects due to temperature changes or do to changes in the cavity size. For more details see Howe (2008) and references therein.

In addition to the frequency-shifts, also all other p-mode characteristics change with the activity cycle (e.g., Salabert et al. 2011), in particular their amplitudes and widths. These changes are believed to be a result of the effect of activity on the excitation and damping of the oscillations. A strong magnetic field can diminish the turbulent velocities in a convectively unstable layer and this can affect the driving of acoustic modes. In solar-type stars, the magnetic field may become sufficiently strong such as to affect not only the properties of the p-mode propagation, but also the turbulence of the convection by reducing its magnitude with increasing stellar activity, thereby reducing the amplitudes of the oscillations. This has been observed as an anti-correlation between Sun spot proxies and the amplitudes of solar-like oscillations (Garc\'\i a et al. 2010). On the other hand the mode widths are correlated with the activity, so the modes have shorter lifetimes in the presence of enhanced activity levels, i.e., this could indicate more changes in the damping rate than in the excitations (Chaplin et al. 2000; Jim\'enez Reyes et al. 2003).

Apart from the frequency variation with a period of 11 years, Fletcher et al. (2010), Broomhall et al. (2011b), and Simoniello et al. (2012) found a quasi biennial periodicity (QBP) of 2-year (right panel of Fig.~\ref{cor}). While this signal could be linked to the 11-year solar cycle, the source is thought to be independent. The properties of the QBP have been interpreted as the visible manifestation of a second dynamo mechanism, which is induced by the strong latitudinal shear located below the surface at 0.95 solar radius. However, the seismic findings do not  rule out  other physical mechanisms proposed so far to explain the origin of the QBP in solar activity proxies such as (i) the instability of magnetic Rossby waves in the tachocline (Zaqarashvili et al. 2010, 2011), (ii) the beating between different dynamo modes (Fluri and Berdyugina 2004)  or merely an amplitude modulation of the dipole component of the dynamo due to non linear effects (Tobias et al. 2005; Moss 2008).


\section{Sun vs stars}
\subsection{Activity effects on oscillations}
\begin{figure}
\includegraphics[width=\linewidth]{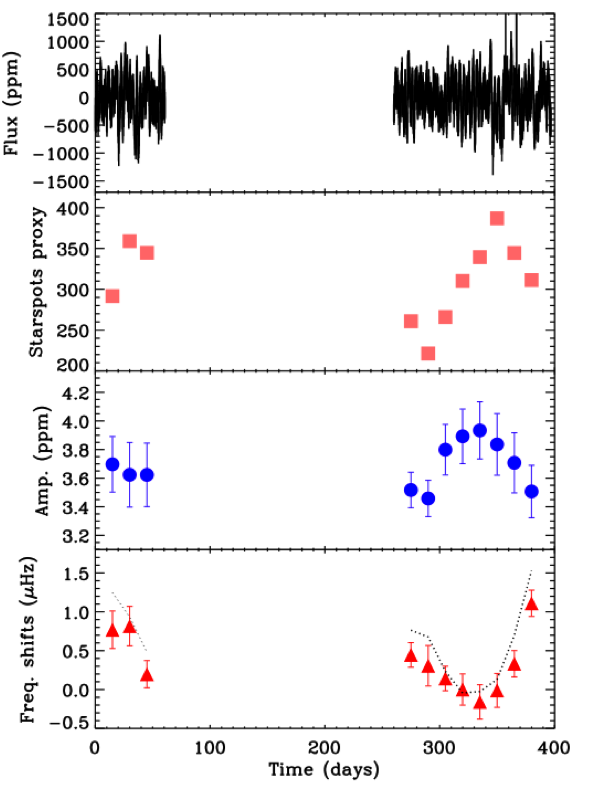}
\caption{Time evolution beginning 6 February 2007 of the mode amplitude (top); the frequency shifts using two different methods (middle), cross correlations (red triangles) and individual frequency shifts (black dots); and starspot proxy (bottom) built by computing the standard deviation of the light curve. All of these were computed by using 30-day-long subseries shifted every 15 days (50\% overlapping). The corresponding 1$\sigma$ uncertainties are shown. Figure taken from Mathur et al. (2011).}
\label{HD49933}
\end{figure}

To place the results of the Sun in context it is important to investigate the connection between surface activity indicators and oscillations in other stars. An important result in this respect was presented by Garc\'\i a et al. (2010). These authors present a magnetic activity cycle in the F5V star HD49933 --observed with CoRoT-- with some characteristics similar to those also found in the solar cycle. These results are shown in Fig.~\ref{HD49933}. The anti-correlations between the oscillation amplitude and frequency shifts on the one hand, and the star spot proxy on the other hand are clear. There is however a phase shift between the variation in the oscillation properties and the activity. The nature of this phase shift is not yet well understood. It could be due to inclination effects  (e.g., V\'azquez Rami\'o et al. 2011), but it is also possible that the oscillations do already feel the magnetic field before it shows itself as spots on the surface as it was the case for the rise of solar cycle 24 (Salabert et al. 2009). Observations of this effect in a larger sample of stars could provide inside in this matter.

Prospects for activity and oscillation measurements  in more stars are promising as first results for an ensemble study of a few hundred solar-like stars observed with \textit{Kepler} have been presented by Chaplin et al. (2011a). Indeed, these authors showed evidence for the impact of stellar activity on the detectability of solar-like oscillations. For these stars the timeseries are not yet long enough to cover a stellar cycle, but by using the activity levels deduced from the variability in the photometric flux they show that the number of stars with detected oscillations drops significantly with increasing levels of activity. So far this is only a statistical study, but with the foreseen length of the \textit{Kepler} mission we expect detailed investigations of individual stars for which one (or more) stellar cycle(s) could be observed in the not-too-distant future.

\subsection{P$_{\rm cycle}$ vs. P$_{\rm rot}$}
As indicated before, rotation plays an important role in our understanding of the dynamos from which the magnetic fields that form starspots originate. This issue has been studied in more detail by e.g., B\"ohm-Vitense (2007) and references therein. They state that the intensification of magnetic fields by a stellar dynamo is due to latitudinal velocity gradients, seen on the surface as differential rotation, created by the interaction between rotation and convection. Additionally, large-scale motions in stars lead to redistribution of angular momentum, which also leads to vertical gradients of rotational velocities. Either of these velocity gradients could be important in the formation of the toroidal field ($\Omega$-effect). B\"ohm-Vitense (2007) suggest that the time it takes to reach the rising point for the toroidal field to form spots determines the length of the activity cycles. Therefore, the relation between the lengths of the rotation periods ($P_{\rm rot}$) and the cycle lengths ($P_{\rm cycl}$) could be of importance to reveal which velocity gradient is of prime importance for stellar activity.

\begin{figure}
\includegraphics[width=\linewidth]{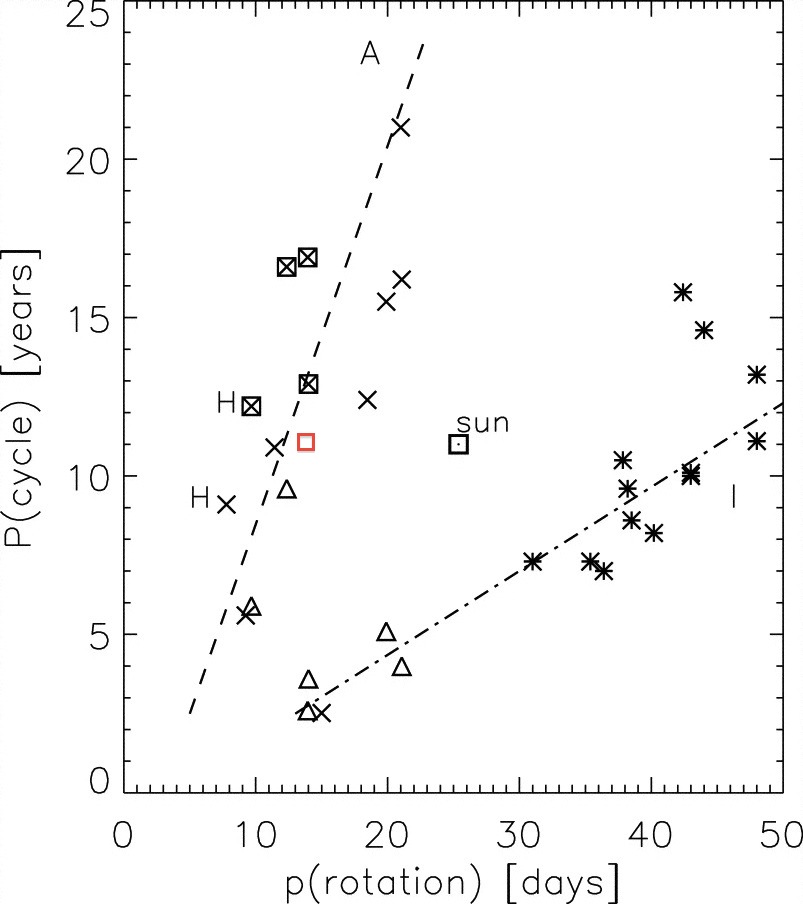}
\caption{Periods of the activity cycles, $P_{\rm cycl}$ in years, are plotted as a function of the rotation periods, $P_{\rm rot}$, in days. The data follow two sequences, the relatively young, active A sequence (dashed line) and the generally older, less active I sequences (dash-dotted line). The letter H indicates Hyades group stars, crosses indicate stars on the A sequence, and asterisks indicate stars on the I sequence. Squares around the crosses show stars with $B - V < 0.62$. Triangles indicate secondary periods for some stars on the A sequence. The solar point is plotted as a square with a dot inside. The red square indicates the Sun at half its known rotational period. Figure adapted from B\"ohm-Vitense (2007).}
\label{rot}
\end{figure}

Measurements of $P_{\rm cycl}$ as a function of $P_{\rm rot}$ show `a distinct segregation of active (A) and inactive (I) stars into two approximately parallel bands'. These results are also shown in Fig.~\ref{rot}. The linear relation for each band indicates that the number of stellar rotations per activity cycle is roughly the same for all stars on a sequence but different for the different sequences. This could indicate that different velocity gradients that feed the dynamos are active in the two sequences. It seems that for the A-sequence stars, which require many rotations per cycle to lift the toroidal field to the surface, differential rotation in relatively thin convection zones near the surface is the dominant driver of the dynamos. The I-sequence stars need fewer rotations per cycle and it seems that large vertical gradients in rotation periods close the the bottom of mixed layers are more important (interface dynamo). The transition between the two regions occurs at a rotational period for which the rotation velocity becomes comparable to the convective velocity. We refer to B\"ohm-Vitense (2007) for more details.

As shown in Fig.~\ref{rot}, the Sun seems to be in the transition between the two sequences and both velocity gradients might be equally important. This gives the Sun a very special position as no other stars are observed in this region. We note here that a reduction of a factor of 2 in the solar rotation period would put the 11-year magnetic cycle on top of the A sequence (see Garc\'\i a 2012 for further details).

With the prospect of measuring solar-like oscillations in a larger number of active stars --thanks to the space instrumentation-- we think it might be possible to infer their internal structures (e.g., Ballot et al. 2004; Mazumdar et al. 2012). In particular the size and properties of the external convective zone could possibly be revealed, and it might also be possible to verify whether internal structure differences between different stars indeed confirm the presence of the two different velocity gradients that act on the magnetic field.

\section{Future}
With the longer timeseries that will become available from the \textit{Kepler} mission, as well as the possibility to revisit some of the already observed targets with CoRoT, it will be possible to obtain activity proxies for many more stars and investigate the relations between those proxies, the oscillations and stellar rotation periods. This will allow us to test whether there are more stars such as the Sun in the $P_{\rm cycl}$ - $P_{\rm rot}$ plane and the stellar structure differences responsible for the different number of stellar rotation cycles needed per activity cycle for the active (A) and less active (I) stars. It will also be possible to investigate how the correlations between the different measures change as a function of stellar parameters, such as metallicity, mass, primoridial rotation and stellar evolution. This could possibly also reveal whether a phase-shift between the phenomena is due to inclination effects or due to influence of the activity on the oscillations prior to the observations of spots.

Although the timeseries provided by \textit{Kepler} are of unprecedented length and accuracy it will still be difficult to study stars with cycles longer than the 11-years one of the Sun. Nevertheless, the observations of stars with shorter cycles (e.g., Garc\'\i a et al. 2010; Metcalfe et al. 2010) will provide a wealth of information to further develop our understanding of the solar / stellar activity and the influence on the seismic properties.

\acknowledgements
SH thanks the organizers of the HelAs V meeting in Obergurgl, Austria for the invitation and the European Science Foundation (ESF) for all arrangements made. SH acknowledges financial support from ESF and the Netherlands Organisation for Scientific Research (NWO).


\begin{thebibliography}{}
\bibitem{} Aerts, C., Christensen-Dalsgaard, J., Kurtz, D.W.: 2010, Springer
\bibitem{} Aizenman, M., Smeyers, P., Weigert, A.: 1977, A\&A~58, 41
\bibitem{} Babcock, H.W.: 1961, ApJ~133, 572
\bibitem{} Ballot, J., Turck-Chi\`eze, S., Garc\'\i a, R.A.: 2004, A\&A~423, 1051
\bibitem{} Basu, S., Antia, H.M.: 1994, MNRAS~269, 1137
\bibitem{} Bedding, T. R.: 2011, in Canary Islands Winter School of Astrophysics, Vol. XXII, Asteroseismology ed. P. L. Pall\'e (Cambridge: Cambridge Univ. Press), in press (arXiv:1107.1723)
\bibitem{} B\"ohm-Vitense, E.: 2007, ApJ~657, 486
\bibitem{} Broomhall, A.-M., Chaplin, W.J., Elsworth, Y., New, R.: 2011a, MNRAS~413, 2978
\bibitem{} Broomhall, A.-M., Fletcher, S.T., Salabert, D., et al.: 2011b, JphCS~271,012025
\bibitem{} Chaplin, W.J., Elsworth, Y., Howe R., et al.: 1996, SoPh 168, 1
\bibitem{} Chaplin, W.J., Elsworth, Y., Isaak, G.R., Miller, B.A., New, R.: 2000, MNRAS~313, 32
\bibitem{} Chaplin, W.J., Bedding, T.R., Bonanno, A., et al.: 2011a, ApJ~732, L5
\bibitem{} Chaplin, W.J., {Kjeldsen}, H. and {Christensen-Dalsgaard}, J., et al.: 2011b, Science~332, 213
\bibitem{} Charbonneau, P., MacGregor, K.B.: 1997, ApJ 486, 502
\bibitem{} Charbonneau, P.: 2010, LRSP~7, 3
\bibitem{} Choudhari, A.R., Sch\"ussler, M., Dikpati, M.: 1995, A\&A 303, L29
\bibitem{} Deheuvels, S., Michel, E.: 2011, A\&A~535, 91
\bibitem{} Denker, C., Goode, P.R., Ren, D., et al.: 2006, SPIE 6267,  62670A
\bibitem{} Dikpati, M.: 2005, ASR~35, 322
\bibitem{} Dikpati, M., Gilman, P.A.: 2007, NJPh~9, 297
\bibitem{} Durney, B.R., Latour, J.: 1978, GApFD 9, 241
\bibitem{} Elsworth, Y., Howe, R., Isaak, G.R., McLeod, C.P., New, R.: 1990, Nature~345, 322
\bibitem{} Fletcher, S.T., Broomhall, A.-M., Salabert, D., et al.: 2010, ApJ~718, L19
\bibitem{} Fluri, D.M., Berdyugina, S.: 2004, Sol. Phys~224, 153
\bibitem{} Garc\'ia, R.A., Pall\'e, P.L., Turck-Chi\`eze, S., et al.: 1998, ApJ~504, L51
\bibitem{} Gar\'\i a, R.A., Turck-Chi\`eze, S., Boumier, P., et al.: 2005, A\&A~442, 385
\bibitem{} Garc\'ia, R.A., Mathur, S., Salabert, D., et al.: 2010, Science~329, 1032
\bibitem{} Garc\'ia, R.A.: 2012, AN, this proceedings
\bibitem{} Hekker, S., Basu, S., Stello, D., et al.; 2011a, A\&A 530, A100
\bibitem{} Hekker, S., Gilliland, R.L., Elsworth, Y., et al.: 2011b, MNRAS 414, 2594
\bibitem{} Hekker, S., Basu, S., Elsworth, Y., Chaplin, W.J.: 2011c, MNRAS 418, L119
\bibitem{} Howe, R.: 2008, ASR~41, 846
\bibitem{} {Jain}, K., {Tripathy}, S.C., {Hill}, F.: 2009, ApJ~695, 1567
\bibitem{} Jefferies, S.M., Pomerantz, M.A., Duvall, T.L., Jr., et al.: 1988, in Seismology of the Sun and Sun-like Stars, ed. E. Rolfe (ESA
SP-286; Noordwijk: ESA), 279
\bibitem{} Jim\'enez, A., Garc{\'{\i}}a, R.A. and {Pall\'e}, P.L.: 2011, ApJ~743, 99
\bibitem{} Jim\'enez Reyes, S.J., {Garc{\'{\i}}a}, R.A., {Jim\'enez}, A., {Chaplin}, W.~J.: 2003, ApJ~595, 446
\bibitem{} Jim\'enez Reyes, S.J., {Chaplin}, W.J., {Elsworth}, Y., {Garc{\'{\i}}a}, R.A.: 2004, ApJ~604, 969
\bibitem{} Judge, P.G., Thompson, M.J.: 2012, IAUS~286, 15 
\bibitem{} Karoff, C.: 2007, MNRAS~381, 1001
\bibitem{} Libbrecht, K.G., Woodard, M.F.: 1990, Nature~345, 779
\bibitem{} Leighton, R.B.: 1964, ApJ 140, 1547
\bibitem{} Leighton, R.B., 1969, ApJ 156, 1
\bibitem{} Mathur, S.,  {Garc{\'{\i}}a}, R.A., {RŽgulo}, C., et al.: 2010, A\&A~511, 46
\bibitem{} Mathur, S.,  {Garc{\'{\i}}a}, R.A., {Salabert}, D., et al.: 2011, JPhCS 271, 012045
\bibitem{} Mazumdar, A., Monteiro, M.J.P.F.G., Ballot, J., et al.: 2012, MNRAS~Submitted
\bibitem{} {Metcalfe}, T.~S., {Basu}, S., {Henry}, T.~J., et al.: 2010, ApJ~723, L213
\bibitem{} Monteiro, M.J.P.F.G., Christensen-Dalsgaard, J., Thompson, M.J.: 1994, A\&A~283, 247
\bibitem{} Moss, D.: 2008, MNRAS~306, 300
\bibitem{} Mosser, B., Belkacem, K., Goupil, M.J., et al.: 2011, A\&A~525, L9
\bibitem{} Osaki, J.: 1975, PASJ~27, 237
\bibitem{} Ossendrijver, M.A.J.H.: 2000, A\&A 359, 1205
\bibitem{} Rempel, M.: JPCS~118, 1
\bibitem{} Schmitt, D.: 1987, ApJ 174, 281
\bibitem{} Schmitt, D., Sch\"ussler M.: 1989, A\&A 223, 343
\bibitem{} {Salabert}, D., {Garc{\'{\i}}a}, R.A., {Pall\'e}, P.L., {Jim\'enez-Reyes}, S.J.: 2009, ApJ~504, L1
\bibitem{} Salabert, D., {Garc{\'{\i}}a}, R.A., {Pall\'e}, P.L., {Jim\'enez}, A.: 2011, JPCS~271, 1
\bibitem{} Simoniello, R., Finsterle, W., Salabert, D., et al.: 2012, A\&A~539, A135
\bibitem{} Tassoul, M.: 1980, ApJS~43, 469
\bibitem{} Tobias, S., Weiss, N., Kirk, N.: 2005, MNRAS~273, 1150
\bibitem{} {V\'azquez Rami\'o}, H., {Mathur}, S., {R\'egulo}, C., {Garc{\'{\i}}a}, R.A.: 2011, JPCS~271, 1
\bibitem{}Vecchio, A., Laurenza, M., Meduri, D., Carbone, V., Storini, M.: 2012, ApJ~749, 27
\bibitem{} Zaqarashvili, T.,  Carbonell,M., Oliver, R., Ballester, J.L.: 2010, ApJL~724, 95
\bibitem{} Zaqarashvili, T., Oliver, R., Ballester, J.L., et al.: 2011, A\&A~532,  139
\bibitem{} Zirin, H.: 1970, S\&T 39, 215
\end{thebibliography}
\end{document}